# Ion behavior and interelectrode breakdown voltage of a drift tube

GENG Hao (耿浩)[1], ZHAO Zhong-Jun (赵忠俊)[2], DUAN Yi-Xiang (段忆翔)*[3]

[1]*Research Center of Analytical Instrumentation, Analytical & Testing Center, Sichuan University, Chengdu 610065, China*

[2]*Research Center of Analytical Instrumentation, College of Chemistry, Sichuan University, Chengdu 610065, China*

[3]*Research Center of Analytical Instrumentation, College of Life Science, Sichuan University, Chengdu 610065, China*

**Abstract:** We experimentally studied ion behavior and interelectrode breakdown voltage. The ion behavior of a drift tube directly influences the detection of ion intensity, and then influences the detection sensitivity of a system. Interelectrode voltage and pressure directly influence the ion behavior. Gas discharge between electrodes influences the adjustments required for interelectrode voltage. The experimental results show: ion intensity increases exponentially with the increment of voltage between drift electrodes; ion intensity decreases exponentially as pressure increases; with the increment of pressure, the breakdown voltage at first decreases, and then increases; ion injection has a significant influence on breakdown voltage, and this influence depends on the pressure and shapes of the electrodes. We explain the results above through assumptions and by mathematical methods.

## 1 Introduction

Drift tubes are used in various technologies, including proton transfer reaction mass spectrometry (PTR-MS), ion mobility spectrometry, selected ion flow drift tube mass spectrometry and injected ion drift tube techniques. These technologies are widely used in atmospheric chemistry, plant studies, food science, medical applications, the detection of chemical warfare agents and the probe of cluster properties [1-9].

Wang *et al.* applied thermal desorption extraction PTR-MS to rapidly determine residual solvent and sterilant measurements. They proposed two novel methods to quantify residual chemicals in solid infusion sets [2]. Yuan *et al.* mounted PTR-MS on an aircraft for atmospheric measurements over the Deepwater Horizon oil spill in the Gulf of Mexico in 2010, and strong signals of cycloalkanes were obtained [3]. Haase *et al.* used PTR-MS to perform research on acetic acid measurements. After calibration, three different configurations of PTR-MS had detection limits from 0.06 to 0.32 ppbv with dwell times of 5s [4]. Agarwal *et al.* used PTR-MS to detect isocyanates and polychlorinated biphenyls. They were able to determine the rapid detection of isocyanates and polychlorinated biphenyls at high accuracy without sample preparation. Results

Supported by financial support from the National Major Scientific Instruments and Equipment Development Special Funds (No.2011YQ030113), the National Recruitment Program of Global Experts (NRPGE), the Hundred Talents Program of Sichuan Province (HTPSP), and the Startup Funding of Sichuan University for setting up the Research Center of Analytical Instrumentation

Corresponding author DUAN Yi-Xiang, E-mail: yduan@scu.edu.cn

for real-time monitoring of industrial waste, polluted air or water quality surveillance were obtained [5]. Liang *et al*. applied a stand-alone ion mobility spectrometer in the detection of black powder. They overcame not only peak overlap but also the negative effect of sulfur ions, and achieved a detection limit of 5 pg [6]. Cheng *et al*. used photoionization ion mobility spectrometry to detect explosives. They used a commercial VUV krypton lamp to ionize acetone of 20 ppm and obtained a stable current of reactant ions of 1.35 nA [7]. Jarrold *et al*. studied reactions of $Si_n^+$ with $C_2H_4$ based on selected ion drift tube techniques [8]. Fhadil *et al*. studied mobilities of various ions of oxygen in the injected-ion drift tube [9].

The structures of these drift tubes are at least similar if not the same, especially in the drift region. Ion behavior in the drift tube directly influences ion intensity detection, and thus influences the detection sensitivity of a system. The ion behavior is controlled by the interelectrode voltage and pressure, as it is difficult to change the drift tube structure. Breakdown voltage is also significant, as the discharge between electrodes influences the adjustment of the interelectrode voltage. Ennis *et al*. and Hanson *et al*. have done some simple studies on the ion behavior of the drift tube [10，11]. To the best of our knowledge, however, we are the first to have systematically studied the ion behavior and interelectrode breakdown voltage of the drift tube. This paper explains not only the exponential variation of ion intensity, but also how ion injection, pressure and electrode shape influences the breakdown voltage. The entire experimental device is shown in fig. 1a. The inner structure of the drift tube and the equipotential lines inside it are shown with the assistance of SIMION software in fig. 1b. SIMION is a powerful software that provides highly interactive and direct ways to simulate electric components [12-14].

## 2 Experimental setup

As shown in fig. 1a, water vapor from a water vapor generator (homemade) was processed through a mass flow controller (Bronkhorst High-Tech B.V) into a silica tube, where the water vapor was ionized by microwave plasma generated by a homemade cavity, with the result that the ions passed through the drift tube (homemade). The power source (customized by Tianjin Dongwen High Voltage Power Supply Co., Ltd) supplied voltage to create an electric field. Being forced by the electric field, the ions moved into an ion detection system. A vacuum system controlled the pressure in the drift tube. The ion detection system is composed of a cylindrical ion detector (homemade) and a current tester (HB-311, Nanjing Hongbin Co., Ltd). Ions impacted the ion detector and the current went through the current tester. Therefore, ion intensity can be expressed as a current of unit nA. The vacuum system is composed of a vacuum pump (Beijing Beiyiyoucheng Vacuum Technology Co., Ltd), a pressure controller (ZDMC-I-LED, Chengdu Zhenghua Electronic Instrument Co., Ltd), and a vacuum gauge (Chengdu Zhenghua Electronic Instrument Co., Ltd). The microwave power (2.45 GHz) was supplied by a solid state microwave generator (Nanjing Yanyou Electronic Science and Technology Co., Ltd). In our experiments, the mass flow controller was set at 1.5sccm, the microwave cavity was tuned to and fixed at an optimized position, and the control voltage for the microwave output power was fixed at 6V. Experimental conditions were set as mentioned above unless otherwise specified. All of the data and results presented from fig.2 to fig.6 were obtained from such an experimental device.

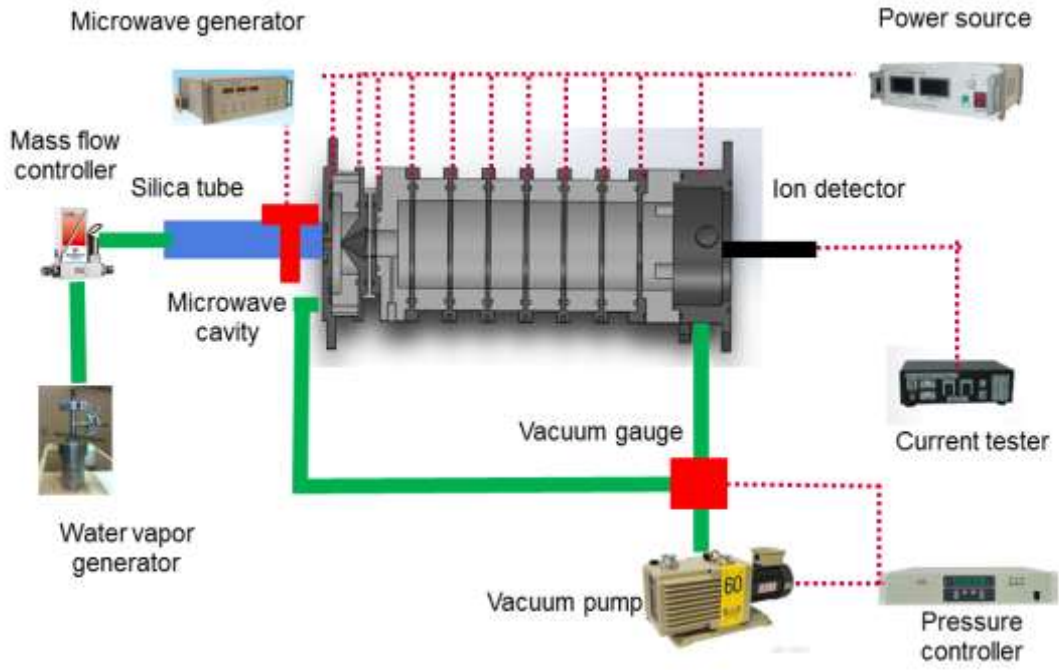

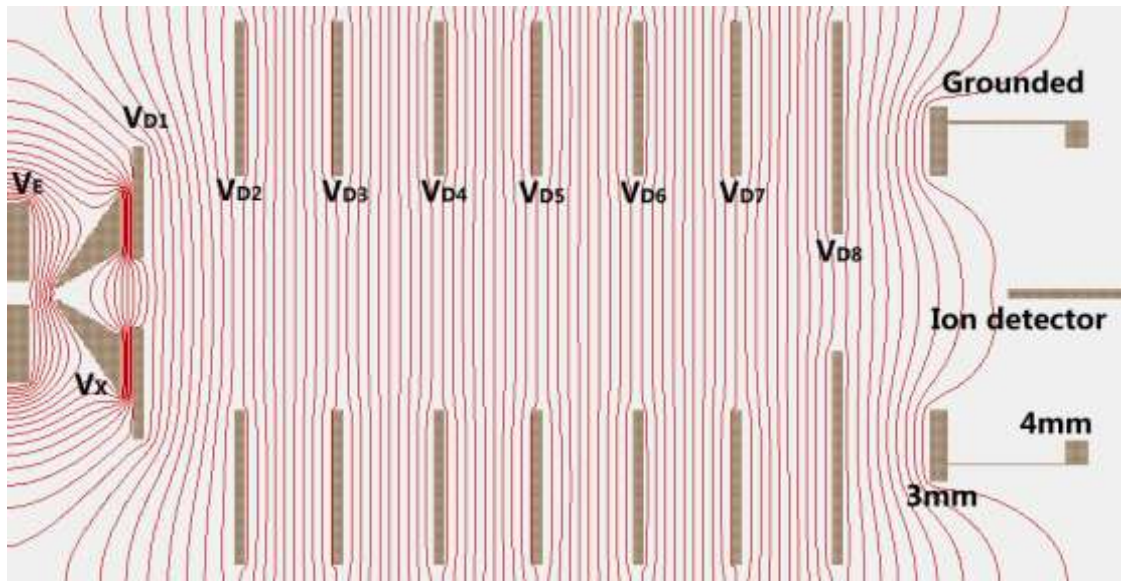

**Fig. 1 (a) The entire experimental device, (b) Inner structure and equipotential lines in the drift tube. (color on line)**

$V_E$ is the potential of the entrance electrode. $V_X$ is the potential of the extraction electrode. With potentials from $V_{D1}$ to $V_{D8}$, the drift electrodes average a total potential $V_D$. For convenience, an electrode potential symbol also stands for its corresponding electrode. Electrodes are isolated from each other by Teflon. $V_E$ is 4mm thick, $V_X$ 11.8mm thick, all the drift electrodes 2mm thick and the grounded electrode 29mm thick. $V_E$ is 4mm in its inner diameter, and $V_X$ 2mm and 13.55mm respectively, $V_{D1}$ 12mm, $V_{D8}$ 20mm, the electrodes from $V_{D2}$ to $V_{D7}$ 40mm and the grounded electrodes 40mm and 50mm respectively. The ion detector is 2mm in diameter. $V_E$ and $V_X$ are 4mm apart, $V_{D1}$ and $V_X$ 2mm, two adjacent drift tubes 15mm, and $V_{D8}$ and the grounded electrode 15mm. The ion detector stretches 14mm into the drift tube. The equipotential lines calculated by SIMION inside the drift tube are shown in the figure, where $V_E$, $V_X$ and $V_D$ are set at 600V, 550V and 500V respectively.

## 3 Results and discussion

By changing voltage and pressure, we measured ion intensity dependence on voltage and pressure, and dependence on pressure of breakdown voltage.

**3.1** Ion intensity dependence on voltage

Ions are driven by the electric field and cause a great number of collisions with particles, the velocity thereof being directly related to the electric field and particle concentration. To describe the relationship between the electric field $E$ and ion velocity $v$, we define ion mobility as [15]:

$$\mu = \frac{v}{E}. \tag{1}$$

If an ion is equal to the particles in the mass, its mobility is:

$$\mu = \frac{q\bar{\lambda}}{2M\bar{v}_t}, \tag{2}$$

Where $\bar{\lambda}$ is the mean free path of the ions, $M$ the mass of the ion and particle, $\bar{v}_t$ the mean thermal motion velocity of the ions, and $q$ the charge on the ion. Otherwise, its mobility is:

$$\mu = \frac{q_0 q \bar{\lambda}}{M \bar{\bar{v}}_t} \sqrt{\frac{M+M_a}{M}}, \tag{3}$$

where $M_a$ is the particle mass, $\bar{\bar{v}}_t$ the root mean square of the ion thermal motion velocity, and $q_0$ a constant. From (2) and (3), we know that the ion mobility $\mu$ is proportional to the mean free path $\bar{\lambda}$. As is well known, $\bar{\lambda}$ is inversely proportional to pressure, and so is the ion mobility.

The ion intensity dependence on the voltage between $V_E$ and $V_X$ at different pressures is shown in fig. 2. $V_X$ is fixed at 300V, and $V_D$ at 200V. With an increment of $V_E$, more ions are extracted into the drift tube. The electric field between $V_E$ and $V_X$ has a deflection effect on the ions, but collisions between particles and ions severely weaken this effect. Overall, ion intensity is enhanced as $V_E$ increases. From equations (2) and (3), ion mobility $\mu$ decreases as the pressure increases. According to (1), it is harder for the electric field to focus the ions. Then, with an increase in $V_E$, the ion intensity obviously increases at low pressure but stays nearly invariant at high pressure.

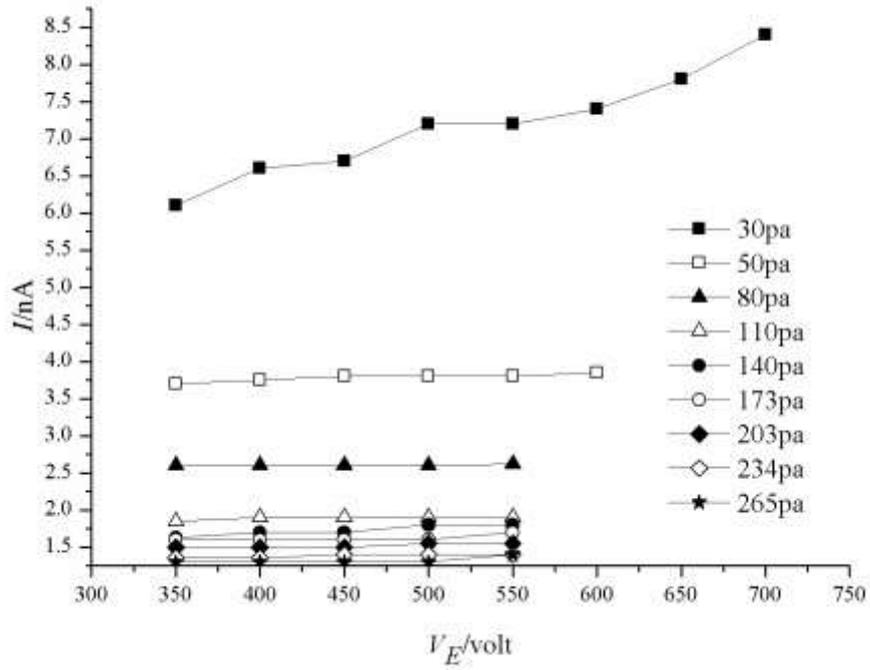

**Fig. 2 Ion intensity dependence on voltage between $V_E$ and $V_X$ at different pressures.**

As shown in fig. 3, ion intensity is enhanced as $V_X$ increases at different pressures. $V_D$ is fixed at 200V, and voltage between $V_E$ and $V_X$ is fixed at 50V. On account of the geometry of electrodes $V_X$ and $V_{D1}$, the electric field between the two electrodes focuses the ion beam. The axial electric field accelerates the ions, and therefore, the ions move less in the radial direction before getting to the ion detector. Then, the ion intensity increases as $V_X$ increases. In consideration of the ion mobility and equations (1-3), it becomes harder for the axial electric field between $V_X$ and $V_{D1}$ to accelerate the ions as pressure increases. In consequence, when $V_X$ increases, ion intensity increases very slightly at relatively high pressure.

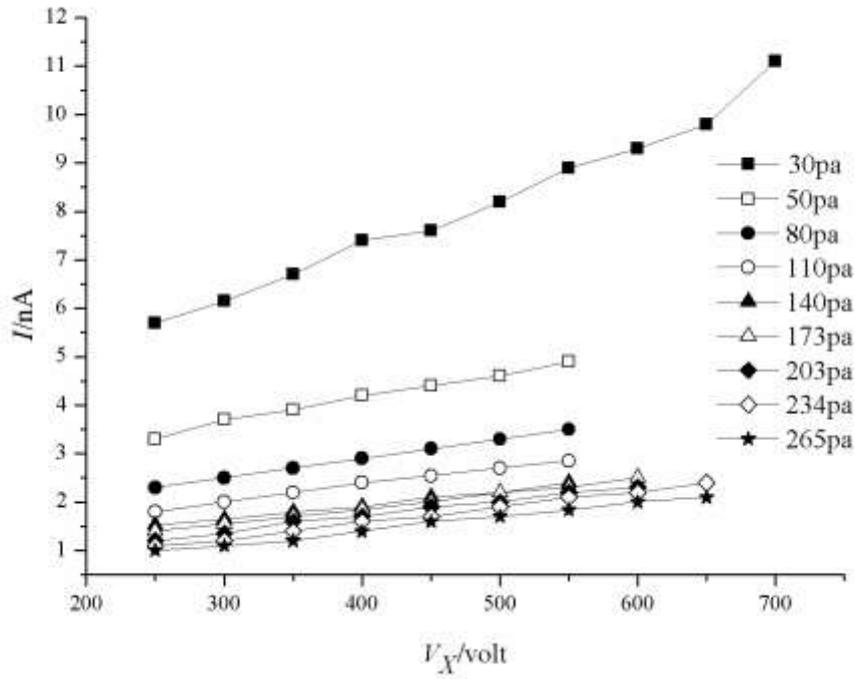

**Fig. 3 Ion intensity dependence on the voltage between $V_X$ and $V_D$ at different pressures.**

Ion intensity dependence on $V_D$ is shown in fig. 4. The exponential curves shown in table 1, where $I$ is the ion intensity, are used to fit the curves in fig. 4.

**Table 1 Fitting curves for experimental data in fig.4.**

| Fitting curve | R-square |
|---|---|
| $I_{30} = 0.22599\exp(0.00829 V_D) + 2.68155$ | 0.99715 |
| $I_{50} = 0.36117\exp(0.00635 V_D) + 1.79274$ | 0.99475 |
| $I_{80} = 0.56549\exp(0.00465 V_D) + 1.45307$ | 0.98938 |
| $I_{110} = 1.0834\exp(0.00337 V_D) + 0.39813$ | 0.99045 |
| $I_{140} = 1.80323\exp(0.00251 V_D) - 1.27219$ | 0.99564 |
| $I_{173} = 1.60344\exp(0.00236 V_D) - 1.26054$ | 0.99681 |
| $I_{203} = 1.86749\exp(0.002 V_D) - 1.27802$ | 0.99335 |
| $I_{234} = 2.69224\exp(0.00153 V_D) - 2.35728$ | 0.99611 |
| $I_{264} = 2.26968\exp(0.00148 V_D) - 1.92219$ | 0.99557 |

As shown in fig.1b, $V_D$ creates an approximately regular parallel uniform electric field. We use a more intuitive method to study the effect of $V_D$ on ion intensity. As $V_D$ increases by $dV_D$, more ions impact the ion detector. With the increment defined as $dn$, obviously it is in positive correlation to the ion number density $n$ and $dV_D$. We propose the hypothesis:

$$\mathrm{d}n = kn\mathrm{d}V_D, \tag{4}$$

where $k$ is the correlation coefficient. Integrating (4) gives:

$$n = n_0 \exp(kV_D), \tag{5}$$

where $n_0$ is the integration constant. Ion intensity is proportional to $n$, so we then deduce:

$$I = A_0 \exp(kV_D) + I_0, \tag{6}$$

where $A_0$ and $I_0$ are both constants. Eq. (6) corresponds with the exponential curves in table 1. From the fitting curves in table 1, we find that parameter $k$ decreases as the pressure increases. With more particles, ions collide more frequently. The resistance, which prevents $V_D$ from focusing ions to impact the ion detector, strengthens as pressure increases. Then, corresponding to the same $\mathrm{d}V_D$, $\mathrm{d}n$ reduces as pressure increases.

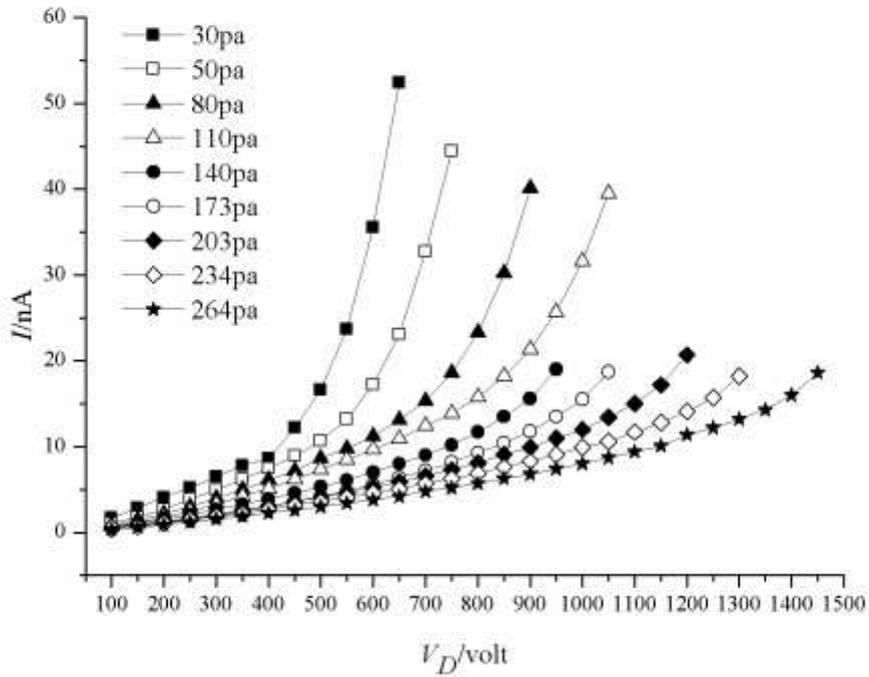

**Fig. 4 Ion intensity dependence on $V_D$ at different pressures.**
The voltage between $V_E$ and $V_X$ was fixed at 50V, and that between $V_X$ and $V_D$ at 10V.

To summarize this section, $V_D$ has much more influence on the ion behavior than other voltages. In an application, $V_D$ should be at high value.

**3.2 Ion intensity dependence on pressure**

Figures 2, 3 and 4 give the visual impression that increases of pressure result in decreases of ion intensity. Dependence of ion intensity on pressure is shown in fig. 5. The curve in fig. 5 is fitted by:

$$I = 14.50274\exp(-0.02777\,P) + 0.76955,\ R\text{-square} = 0.98174,$$

where $I$ is the ion intensity, and $P$ the pressure. Assume that within a unit time interval, an ion collides $n_1$ times on average. Define the mean velocity of the ions as $\bar{v}$, then an ion collides $n_1/\bar{v}$ times per unit distance. Define the collision cross section as $\sigma$, it indicates the probability for an ion to collide with particles. Therefore, if an ion moves a distance d$s$, its collision frequency is [15-17]:

$$N\sigma \mathrm{d}s = (n_1/\bar{v})\ \mathrm{d}s, \tag{7}$$

where $N$, the number density of particles, is proportional to pressure. Define $n$ as the number of ions that move a distance $s$ without collision. The number of ions that collide at a distance between $s$ and $s+\mathrm{d}s$ is:

$$\mathrm{d}n = -(n_1/\bar{v})n\mathrm{d}s = -N\sigma n\ \mathrm{d}s, \tag{8}$$

where the minus sign in equation (8) indicates a reduction of ions. Integrating (8) gives:

$$n = n_0\exp(-N\sigma s), \tag{9}$$

where $n_0$ is the integration constant. The ion intensity is proportional to the number of ions reaching the ion detector. Parameter $\sigma$ depends on the radius of the particles, and $s$ is the geometric size of the drift tube. They are constants, so the ion intensity can be expressed as an exponential function of pressure $P$:

$$I = A_1\exp(-B_0 P) + I_1, \tag{10}$$

where $A_1$, $I_1$ and $B_0$ are undetermined constants. The ion intensity decreases exponentially with the increases of pressure. Assigning the constants in (10) with the values from the fitting curve of fig.5 ($A_1 = 14.50274$, $B_0 = 0.02777$ and $I_1 = 0.76955$), the derivative of (10) is then:

$$I'(P) = -0.4027410898\exp(-0.02777 P). \tag{11}$$

The absolute value of $I'(P)$ is lower than $1\times 10^{-4}$(nA/pa) when $P$ is higher than 300pa, $|I'(300)| = 9.70328106475187\times 10^{-5}$(nA/pa). According to fig.5, when the pressure is lower than 300pa, $|I'(P)|$ is big enough to make the ion intensity increase notably; when the pressure is higher than 300pa, $|I'(P)|$ is too small to obviously influence the ion intensity.

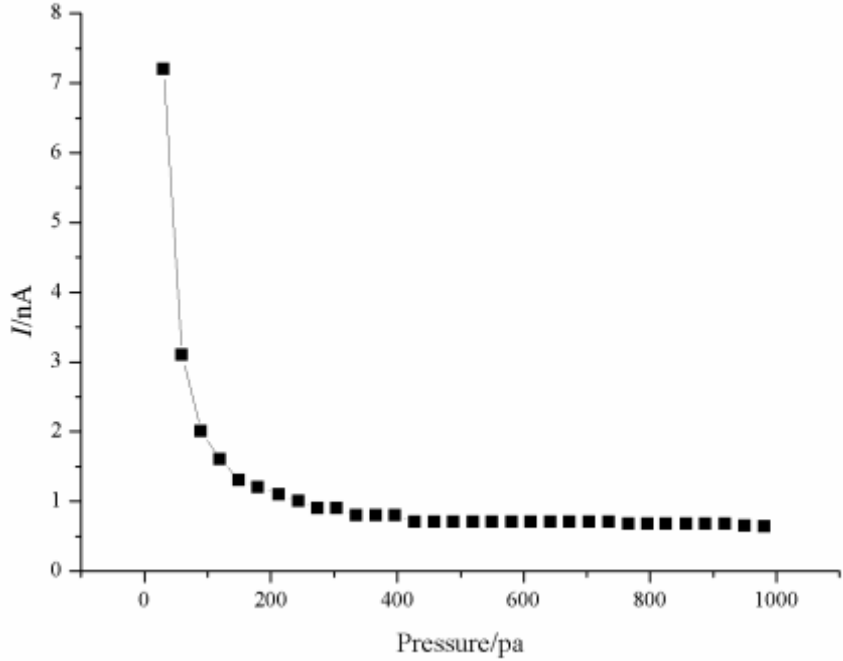

**Fig. 5 Ion intensity dependence on pressure.**
$V_E$, $V_X$ and $V_D$ were fixed at 400V, 300V and 200V.

**3.3** Breakdown voltage.

By adjusting the potentials of $V_E$ and $V_X$, we find that when gas breakdown occurs at a potential larger than a certain value, a self-maintained discharge occurs between the electrodes. An ion is very unlikely to have energy as high as the thousands of volts needed to ionize neutral particles through collisions. Consequently, ionization through collisions between neutral particles and ions is neglected. Generally, ionization of neutral particles is induced by electrons generating an avalanche breakdown (how an avalanche breakdown is caused is shown in Appendix A). Electrons are produced during collisions between ions and electrodes. Based on this knowledge, the discharge formula is [15, 17]:

$$\frac{1}{\gamma} = \exp(\alpha d) - 1, \qquad (12)$$

where $\alpha$ is the first Thompson Ionization Coefficient, which describes the mean number of particles an electron ionizes per unit distance, $d$ is the distance between electrodes, and $\gamma$ is the third Thompson Ionization Coefficient, which describes the mean number of electrons that each ion produces when it impacts an electrode. With the mean free path of electrons defined as $\overline{\lambda_e}$, the probability is $\exp(-x/\overline{\lambda_e})$ that an electron will have a free path bigger than $x$. Therefore, the probability is $\exp(-V_i/E\overline{\lambda_e})$ for an electron to acquire energy $eEx$ which is not less than the ionization energy $eV_i$, where $E$ is the electric field. Therefore, the probability for an electron to ionize particles in distance $\overline{\lambda_e}$ is $\exp(-V_i/E\overline{\lambda_e})$. We can then deduce:

$$\alpha = \frac{1}{\overline{\lambda_e}}\exp(-V_i / E\overline{\lambda_e}). \tag{13}$$

$\overline{\lambda_e}$ is inversely proportional to pressure $P$:

$$\frac{1}{\overline{\lambda_e}} = AP, \tag{14}$$

$$B = V_i A, \tag{15}$$

where $A$ and $B$ are constants. Then we have:

$$\alpha = AP\exp(-BP/E). \tag{16}$$

The logarithmic form of (12) is:

$$\ln(\frac{1}{\gamma} + 1) = \alpha d. \tag{17}$$

Substituting $E = \frac{V_b}{d}$ into (16), where $V_b$ is the breakdown voltage, gives:

$$\alpha = AP\exp(-BPd/V_b). \tag{18}$$

Substituting (18) into (17) gives:

$$\ln(\frac{1}{\gamma} + 1) = APd\exp(-BPd/V_b). \tag{19}$$

The logarithmic form of (19) is:

$$\ln(\frac{1}{APd} \ln(\frac{1}{\gamma} + 1)) = -BPd/V_b. \tag{20}$$

The breakdown voltage $V_b$ can then be expressed as:

$$V_b = BPd / \ln\left(\frac{APd}{\ln(\frac{1}{\gamma} + 1)}\right). \tag{21}$$

The experimental data of breakdown voltage dependence on pressure are shown in fig. 6.

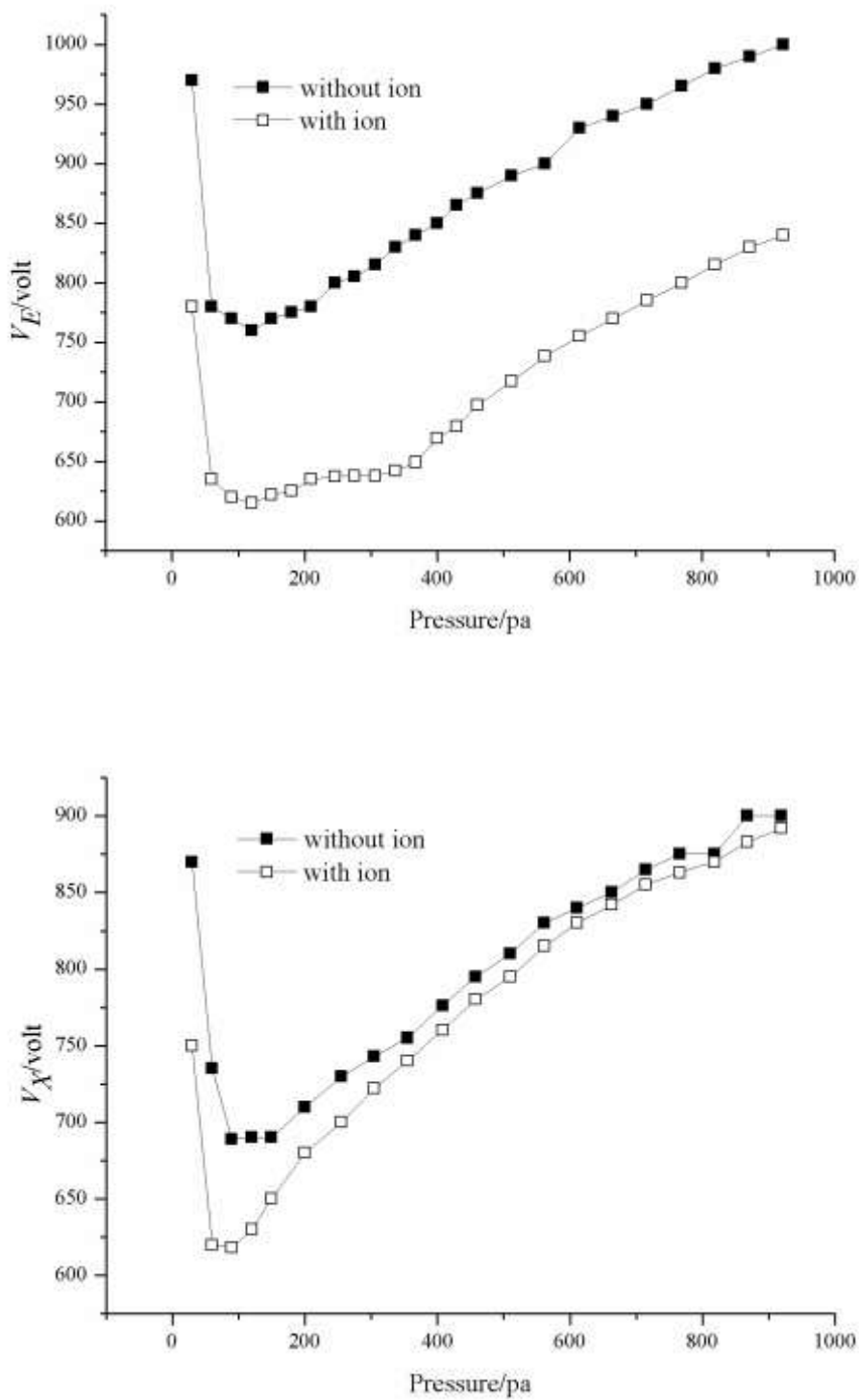

**Fig. 6 Dependence on pressure of breakdown voltage between (a) $V_E$ and $V_X$, (b) $V_X$ and $V_D$.**

$V_D$ is fixed at 200V, $V_X$ at 300V. Labels "with ion" and "without ion" indicate whether the ions are injected in or not.

$V_D$ was fixed at 200V, the voltage between $V_X$ and $V_E$ at 50V. Labels "with ion" and "without ion" indicate whether the ions are injected in or not.

Obviously, ion injection makes discharge easier, as shown in fig. 6. In ion injection, more ions impact the electrodes to create electrons. Accelerated by an electric field, the electrons bombard residual gas and generate secondary electrons, and then all the electrons repeat the process described above, thus causing an avalanche breakdown. As pressure increases, *i.e.* the number of particles increases, the ions will collide more frequently when moving. Therefore, it is harder for ions to acquire enough energy to impact the electrodes and generate electrons. Therefore, little ion injection effect on breakdown voltage between $V_X$ and $V_D$ is observed in fig. 6b as pressure increases. Electrodes with small radii of curvature distort the electric field and the electric field is non-uniform in distribution between the electrodes. This leads to a weakened electric field in some regions, but it sharply enhances the local electric field near the cusp of the electrode. This strong electric field offsets much resistance from particles in the ion movement to the degree that ions still get enough energy to impact electrodes to produce electrons. Consequently, injected ions still have significant influence on the breakdown voltage between $V_E$ and $V_X$ in fig. 6a as pressure increases.

The discussions above are based on reality. However, we could also use a fictitious model to explain the ion injection effect on the breakdown voltage. In the drift tube, besides injected ions, there are also inherent ions. Assume that the inherent ions in the drift tube impact the electrodes and produce $m$ electrons; afterwards, more ions are injected in and the total number of electrons increases to $n(n>m)$. The process above is real, but we now follow with a fictitious model where no more ions are injected in, but $\gamma$ of the inherent ions are amplified by $n/m$. According to the definition of $\gamma$, the number of electrons also increases equivalently from $m$ to $n$. From eq. (21), the increment of $\gamma$ leads to the reduction of $V_b$. Nevertheless, the injected ions must contribute to produce electrons. If this is not so, then the fictitious increment of $\gamma$ is unreasonable. With reference to the analysis of how the electrode's shape influences the electric field in the previous paragraph, when the pressure increases, the fictitious increment of $\gamma$ is still reasonable between $V_E$ and $V_X$ but unreasonable between $V_D$ and $V_X$. Then, the $V_b$ difference between "with ion" and "without ion" in fig.6a is much larger than that in fig.6b as pressure increases.

The subuliform hole of the $V_X$ electrode exaggerates the injection efficiency of the ions; but on the other hand, it also makes the breakdown voltage lower and narrows the adjustable range of voltage between electrodes. Therefore we should weigh between the injection efficiency and adjustable range of voltage when designing drift tubes.

In view of section 3.2 and 3.3, at 30pa, the ion intensity is at the largest value and the breakdown voltage is not at the lowest value. Consulting section 3.1, at 30pa, set gap between $V_E$ and $V_X$ at 400V, gap between $V_X$ and $V_D$ at 500V, and $V_D$ at 650V ($V_E$=1550V, $V_X$=1150V, $V_D$=650V), we then have intensive ion intensity of 63.8nA.

## 4  Conclusions

In this paper, we studied ion behavior and breakdown voltage in a drift tube experimentally. Experiments and theoretical analysis presented the conclusions deduced below:

1. Ion intensity was enhanced as interelectrode voltage increased. Ion intensity grew exponentially with the increment of $V_D$. This phenomenon was explained by an assumption and subsequent mathematical derivation. In practical applications, voltages between electrodes could be adjusted according to actual requirements.

2. Ion intensity decreased exponentially as pressure increased. Formulas based on the collision cross section were used to explain it. The fitting curve of the experimental data matched the derived formulas.

3. The breakdown voltages between electrodes decreased at first and then increased with the increments of pressure. Experimental results showed that the injection of ions made gas breakdown easier, and this effect was also related to the pressure and electrode shapes. The ion injection effect on the breakdown voltage is explained based on the discharge formula.

The results above guide us not only to achieve high detection sensitivity, but also to a suitable drift tube design.

**Appendix A**

Accelerated by the electric field, $n$ electrons move through distance $s$. After particle ionization, the number of electrons has increased by $dn$. $\alpha$ is the first Thompson Ionization Coefficient, and then we have:

$$dn = n\alpha ds. \tag{1A}$$

Integrate (1A) and we have:

$$n = n_0 \exp(\alpha s), \tag{2A}$$

where, $n_0$ is a constant. Therefore, the number of electrons $n$ increases exponentially with distance $s$ and an avalanche breakdown is caused.